\documentclass[12pt]{article}
\usepackage[dvips]{epsfig}

\begin{document}

\title{Vacuumless kinks systems from vacuum ones, an example}
\author{A. de Souza Dutra$^{a,b}$\thanks{%
E-mail: dutra@feg.unesp.br} and A. C. Amaro de Faria Jr.$^{b}$ \\
$^{a}$Abdus Salam ICTP, Strada Costiera 11, 34014 \, Trieste \, Italy\\
$^{b}$UNESP-Campus de Guaratinguet\'{a}-DFQ\thanks{
Permanent Institution}\\
Av. Dr. Ariberto Pereira da Cunha, 333\\
C.P. 205\\
12516-410 Guaratinguet\'{a} SP Brasil}
\maketitle

\begin{abstract}
Some years ago, Cho and Vilenkin, introduced a model which
presents topological solutions, despite not having degenerate
vacua as is usually expected. Here we present a new model with
topological defects, connecting degenerate vacua but which in a
certain limit recovers precisely the one proposed originally by
Cho and Vilenkin. In other words, we found a kind of ``parent''
model for the so called vacuumless model. Then the idea is
extended to a model recently introduced by Bazeia et al. Finally,
we trace some comments the case of the Liouvlle model.
\newline

\end{abstract}

\newpage

Usually the topological objects like domain walls, strings and
monopoles appears when the models support at least two degenerate
vacua. Notwithstanding, there are some models which defy this
commonsense, like the Liouville model \cite{jackiw} -
 \cite{menotti}, the vacuumless (VL) model
introduced originally by Cho and Vilenkin \cite{cho} - \cite{sen}
and, more recently a model where the kink interpolates between two
inflection points instead vacua \cite{deformed}. Here we are going
to concentrate our attention to the VC case, which was originally
studied regarding gravitational aspects of the topological defect
\cite{cho}, and then regarding its topological properties
\cite{dionisio99}, and after that make a discussion in general
lines about how to implement a similar procedure in the other two
cited cases. The Lagrangian density of the model we are going to
introduce here is the usual one for a scalar field,

\begin{equation}
\mathcal{L}=\frac{1}{2}\partial_\mu \varphi \partial^\mu \varphi \, - \,
V\left( \varphi \right)
\end{equation}

\noindent where the potential is given by

\begin{equation}
V\left( \varphi \right) =\frac{1}{2}\left( Acosh(\varphi )-Bsech(\varphi
)\right) ^{2}.  \label{1}
\end{equation}

Note that, if $A=0$ and $B=\pm \,\mu $, we recover the usual vacuumless
potential \cite{dionisio99}

\begin{equation}
V\left( \varphi \right) =\frac{\mu ^{2}}{2}\left( sech(\varphi )\right) ^{2}.
\end{equation}

\noindent Let us use the BPS approach \cite{BPS}, in order to
present the solution of this and the new model we introduced
above. For this, one can write the potential in terms of the so
called superpotential, which is given by
\begin{equation}
\,V\left( \phi \right) =\frac{1}{2}\,W_{\phi }^{2},
\end{equation}

\noindent from which the energy of the static configuration can be obtained
as

\begin{equation}
E_{BPS}=\frac{1}{2}\int_{-\infty }^{\infty }dx\left[ \left( \frac{d\phi }{dx}%
-W_{\phi }\right) ^{2}+W_{\phi }\frac{d\phi }{dx}\right] .
\end{equation}

Observing this equation, we note that the field configuration
which minimizes the energy will obeys the first-order differential
equation
\begin{equation}
\frac{d\phi }{dx}=W_{\phi }\left( \phi \right) ,
\end{equation}
\noindent and his energy is written as
\begin{equation}
E_{BPS}=|W\left( \phi \left( \infty \right) \right) -W\left( \phi \left(
-\infty \right) \right) |.
\end{equation}

Let us now apply this machinery to the above mentioned models. In the case
of the VL model, one can check that the superpotential is given by \cite
{dionisio99}
\begin{equation}
W\left( \phi \right) =2\,B\,\tan ^{-1}\left[ \tan \left( \frac{\phi }{2}%
\right) \right] ,
\end{equation}

\noindent and its slowly divergent kink looks like
\begin{equation}
\phi \left( x\right) =\sinh ^{-1}\left( B\,\,x\right) .  \label{2}
\end{equation}

On the other hand, in the case of the model which we are introducing here (%
\ref{1}), the superpotential has the appearance
\begin{equation}
W\left( \phi \right) =A\,\sinh \left( \phi \right) -2\,B\,\tan ^{-1}\left[
\tan \left( \frac{\phi }{2}\right) \right] ,  \label{superpot}
\end{equation}

\noindent and the corresponding kink and antikink are expressed as
\begin{equation}
\phi \left( x\right) =\pm \,\sinh ^{-1}\left[ \sqrt{\frac{\left( A-B\right)
}{A}}\,\tan \left( \sqrt{A\left( A-B\right) }\,\,x\right) \right] ,
\label{3}
\end{equation}

\noindent from which it can be verified that the expected limit
(\ref{2}) when $A\rightarrow 0$ is really achieved. It is possible
to observe too, from the Figure 1, that when $A\rightarrow 0$ the
vacua of the model becomes more and more far from each other, in
such a way that one can think the VL model \cite{cho}, as a limit
of this model with usual degenerate vacua. Once in this case, the
limit of the field configuration at $x\rightarrow \pm \,\infty $,
are the vacua of the model, we can assure that in these limits,
the field given in (\ref{3}) goes to
\begin{equation}
\phi \left( \pm \infty \right) =\pm \,\cosh ^{-1}\left( \sqrt{\frac{B}{A}}%
\right) .
\end{equation}

\noindent Note that, for consistency, the model will have two
minima provided that $B>A$, otherwise the potential has only one
minimum and the solution of the equation (\ref{3}) presents
singularities at finite points in the space, so rendering itself
as a nonphysical solution. The BPS energy of this configuration
will then be given by
\begin{eqnarray}
E_{BPS}\left( A,B\right) &=&2\,\left| \left( A+\sqrt{A\,B}\right) \sqrt{%
\frac{A+B-2\sqrt{A\,B}}{B-A}+}\right.  \nonumber \\
&& \\
&&\left. -2\,B\,\tan ^{-1}\left[ \tanh \left( \frac{1}{2}\cosh ^{-1}\left(
\sqrt{\frac{B}{A}}\right) \right) \right] \right| .  \nonumber
\end{eqnarray}

Let us now analyze the limit of this energy when $A\rightarrow 0$.
The first term vanishes obviously, and in the second we see that
the argument of the function diverges and, as we know, the inverse
function of the hyperbolic cosine diverges too, but the hyperbolic
tangent of infinity is simply one. As a consequence we conclude
that the limit of the above energy is simply given by
\begin{equation}
E_{BPS}\left( 0,B\right) =\,\pi \,B,
\end{equation}

\noindent which is in absolute accordance with the expected for
the VL model \cite{dionisio99}. The energy density of the model we
are studying is
\begin{equation}
\varepsilon \left( x\right) =\frac{\left( A-B\right) ^{2}\sec \left( \sqrt{%
A\left( A-B\right) \,x}\right) ^{4}}{\left( 1+(1-B/A)\tan \left( \sqrt{%
A\left( A-B\right) \,x}\right) ^{2}\right) },
\end{equation}
\noindent and, as expected, have the correct limit when
$A\rightarrow 0$, becoming itself equal to that of the VL model
\begin{equation}
\varepsilon \left( x\right) =\frac{B^{2}}{\left( 1+B^{2}x^{2}\right) },
\end{equation}
\noindent but for a fixed value of the parameter $B$, the VL model
have a bigger and less concentrated energy density, as can be seen
in the Figure 2. Now, we can discuss the linear stability of the
model here presented. In fact, as shown in \cite{bazeia2} for the
case of coupled scalar fields, the linear stability of the model
with one scalar field can be done as usual by performing small
perturbations on the kink solution,
\begin{equation}
\phi \left( x,t\right) =\phi _{kink}\left( x\right) +\eta \left( x,t\right) .
\end{equation}

\noindent Taking into account only up to the first order terms in
the perturbation, which leads to a Schroedinger-like equation for
the perturbation field
\begin{equation}
\left( -\frac{d^{2}}{dx^{2}}+V_{\phi \phi }\left( \phi \equiv \phi
_{kink}\left( x\right) \right) \right) \eta _{n}\left( x\right) =\omega
_{n}^{2}\,\eta _{n}\left( x\right) ,
\end{equation}

\noindent where $\eta \left( x,t\right) \equiv \sum_{n}\,\eta _{n}\left(
x\right) \,\cos \left( \omega _{n}\,t\right) $.  It is not difficult to see
that the above equation can be achieved from the following ladder operators,
\begin{equation}
a_{\pm }\equiv \pm \,\frac{d}{dx}+W_{\phi \phi },
\end{equation}

\noindent whose Hamiltonian operator $\hat{H}=a_{+}\,a_{-}$, as
shown in \cite{bazeiaPLA96} for general coupled real scalar
fields, have their eigenvalues positive definite and, as a
consequence, the models are stable under small quantum
fluctuations.

In our case, the potential to which the small fluctuations feel,
once again has the VL one as its limit, coming from above as can
be observed in Figure 3.

The bosonic ground stated, which is granted by the translational invariance
in this case, in general can be obtained through the solution of the
equation,
\begin{equation}
\left( -\frac{d}{dx}+W_{\phi \phi }\right) \psi _{0}\left( x\right) =0,
\end{equation}

\noindent where, as a simplified notation, from now on we define that $%
W_{\phi \phi }\equiv W_{\phi \phi }\left( \phi \equiv \phi _{kink}\right) $.
Here we note that, one can rewrite the above equations as
\begin{equation}
\frac{d\psi _{0}\left( x\right) }{\psi _{0}\left( x\right) }=W_{\phi \phi
}\,dx,
\end{equation}

\noindent but we know from the BPS equation that $dx=\frac{d\phi
}{W_{\phi }} $, in such a way that a direct relation between the
bosonic zero-mode and the superpotential can be obtained,
\begin{equation}
\psi _{0}\left( x\right) =N_{0}\,W_{\phi }=N_{0}\,\frac{\sqrt{A}\left(
A-B\right) \left( \sec \left( \sqrt{A\left( A-B\right) }\,x\right) \right)
^{2}}{\sqrt{1+\left( A-B\right) \left( \tan \left( \sqrt{A\left( A-B\right) }%
\,x\right) \right) ^{2}}},
\end{equation}

\noindent where $N_{0}$ is the normalization constant, and his shape is
quite similar to that of the VL model. This allow us to show that the
normalization of the zero-mode is related to the BPS energy through
\begin{equation}
\int \left| \psi _{0}\left( x\right) \right| ^{2}dx=N_{0}^{2}\int \,W_{\phi
}^{2}\,dx=N_{0}^{2}\int \,W_{\phi }\,d\phi =N_{0}^{2}\,E_{BPS}\equiv 1,
\end{equation}
\noindent and we get finally the normalized bosonic zero-mode
\begin{equation}
\psi _{0}\left( x\right) =\sqrt{\frac{1}{E_{BPS}}}\,W_{\phi },
\end{equation}

\noindent apart from an arbitrary constant phase factor. Let us now try to
calculate the fermionic zero-mode. Using in this case, as done by Bazeia in
\cite{dionisio99}, the Yukawa coupling giving by $f\left( \phi \right) \,%
\bar{\psi}\,\psi $, where it is chosen $f\left( \phi \right) =g\,W_{\phi
\phi }$ ($g=1$, in order to get a supersymmetric version of the model \cite
{shiffman}), we reach the following equation for Dirac fermions
\begin{equation}
i\,\gamma ^{1}\frac{d\Psi }{dx}+f\left( \phi \right) \,\Psi =0,\,\Psi
=\left(
\begin{array}{l}
\psi _{+} \\
\psi _{-}
\end{array}
\right) ,
\end{equation}

\noindent and using the representation where $i\,\gamma ^{1}\rightarrow
\sigma _{3}$ we obtain the following equations for the spinor components,
\begin{equation}
\pm \,\frac{d\psi _{\pm }}{dx}+f\left( \phi \right) \psi _{\pm }=0.\,
\end{equation}

\noindent The above equations can be expressed as
\begin{equation}
\frac{d\psi _{\pm }}{\psi _{\pm }}=\mp f\left( \phi \right) \,dx=\mp
g\,W_{\phi \phi }\,\frac{d\phi }{W_{\phi }},
\end{equation}

\noindent which integration gives us finally the spinor
\begin{equation}
\Psi =\left(
\begin{array}{l}
C_{+}\,\,W_{\phi }^{-g} \\
C_{-}\,\,W_{\phi }^{g}
\end{array}
\right) ,
\end{equation}

\noindent where $C_{\pm }$ are arbitrary integration constants. However,
supposing that the function $W_{\phi }$ is well-behaved, vanishing when $%
x\rightarrow \pm \,\infty $. The normalization of the above spinor,
\begin{equation}
\int \left| \Psi \right| ^{2}dx=\int dx\,\left[ \left| C_{+}\right|
^{2}\,\,W_{\phi }^{-2\,g}+\left| C_{-}\right| ^{2}\,\,W_{\phi
}^{2\,g}\right] \equiv 1,
\end{equation}

\noindent will impose that one of the above arbitrary constants must be
chosen equal to zero. Otherwise, the spinor will be not square integrable
and, as a consequence, we are left with two possible solutions, depending on
the signal of $g$,
\begin{equation}
\Psi _{+}=C_{+}\,\,W_{\phi }^{-g}\left(
\begin{array}{l}
1 \\
0
\end{array}
\right) ,\,g<0;\,\Psi _{\_}=C_{-}\,\,W_{\phi }^{g}\left(
\begin{array}{l}
0 \\
1
\end{array}
\right) ,\,g>0.
\end{equation}

In fact, the normalizability of the spinor, implies into further conditions
over the constant $g$. Let us return to the normalization integration, now
simply given by
\begin{equation}
\int \left| \Psi _{\pm }\right| ^{2}dx=\left| C_{\pm }\right| ^{2}\,\int
dx\,\,W_{\phi }^{\left( \mp 2\,g\right) }=\left| C_{\pm }\right| ^{2}\int
\,\,W_{\phi }^{\left( \mp 2\,g\,-1\right) }d\phi .
\end{equation}

Once again, we note that $\left| g\right| \leq \frac{1}{2}$ or the
integration may diverge. At this point, however, some differences can appear
depending which model is being considered. In order to be quite clear on
this point, let us take for instance the limiting case $g=\pm \frac{1}{2}$,
where we have
\begin{equation}
\left| C_{\pm }\right| ^{2}\int \,\,W_{\phi }^{\left( \mp 2\,g\,-1\right)
}d\phi =\left| C_{\pm }\right| ^{2}\int \,d\phi =\left| C_{\pm }\right|
^{2}\left( \phi \left( +\infty \right) -\phi \left( \infty \right) \right) .
\end{equation}

\noindent It is evident that in models like the VL, the zero-mode
fermion can not be normalizable, due to the divergence of the kink
profile \cite{dionisio99}. However, for any usual topological
model with different finite vacua, this case is absolutely
admissible. So, the model we have proposed in this work, can have
its fermionic zero-mode well defined for any value of the
parameter $A$ different of zero, when it becomes equivalent to the
VL model. So, we can think this model as a kind of regularizing
potential, where one can make the vacua arbitrarily far from each
other, without losing the finiteness characteristic of the usual
BPS kinks. In fact, in the VL limit, the normalization constant
tends to zero and the zero mode wave function vanishes. Only in
the VL limit the $g=\pm 1/2$ must be avoided.

Now, considering the cases where a supersymmetric extension of the model is
allowed \cite{shiffman}, $g=\pm 1$. The normalization of the fermionic
zero-mode, becomes quite similar to that of its bosonic counterpart,
\begin{equation}
C_{\pm }=\sqrt{\frac{1}{E_{BPS}}}.
\end{equation}

Let us now briefly discuss the extension of this idea to other
unusual kink models. For instance we take as our next example, the
model introduced recently by Bazeia, Losano and Malbouisson (BLM)
\cite{deformed}. This model is unbounded from below and presents
no vacua, just a maximum at the origin. Notwithstanding, it has a
kink (and antikink) connecting their two inflection points.
Concretely, his potential is given by
\begin{equation}
V_{BLM}\left( \phi \right) =\frac{1}{2}\left( 1-\phi ^{2}\right) ^{3}.
\end{equation}

Following the idea above introduced, we now propose an alternative
model which presents two local minima. In fact it is not yet a
standard one, once it is also unbounded from bellow. For this case
the ``parent'' model is defined as
\begin{equation}
V_{P}\left( \phi \right) =\frac{1}{2}\left( 1-\phi ^{2}\right) \left(
A-1+\phi ^{2}\right) ^{2}.
\end{equation}

\noindent Both potentials are plotted in Figure 4. Once again,
when one takes the limit $A\rightarrow 0$, the BLM model is
recovered. Again there is no problem in consider $g=\pm
\frac{1}{2}$, when calculating the fermionic zero mode. The kink
solution of the parent model in this case is given by
\begin{equation}
\phi \left( x\right) =\pm \,\tanh \left\{ \sinh ^{-1}\left[ \sqrt{\frac{%
\left( A-1\right) }{A}}\,\tan \left( \sqrt{A\left( A-1\right) }\,\,x\right)
\right] \right\} .
\end{equation}

\noindent In fact this is nothing but the deformation of the first
model proposed here, precisely in the same way in which the BLM\
model can be viewed as a deformation of the VC model
\cite{deformed}.

Finally, let us make a brief comment about the case of the
Liouville model, where
\begin{equation}
V\left( \phi \right) =\left( \frac{m}{\beta }\right) ^{2}e^{\beta \,\phi },
\end{equation}

\noindent which evidently does not presents any vacuum. It is
possible to create a model composing a series of exponential
factors, in such a way that we could have a ``parent'' model also
here. However, at least this simple extension is not exactly
solvable. In this case only a numerical solution is available in
principle, and we are not going to consider it.

Our last comment in this work, is that we think that it is
possible to find orthodox ``parent'' kink models for those which
unorthodox features, like the VL, BLM and Liouville models. At
least when is is possible, as shown in the case of the VL model
here, we get a kind of regularization of the kink properties,
softening some of his properties.

\vfill

\noindent \textbf{Acknowledgments:} The authors are grateful to CNPq and
CAPES for partial financial support, and ASD to the Professor D. Bazeia for
introducing him to this matter. This work has been finished during a visit
(ASD)\ within the Associate Scheme of the Abdus Salam ICTP.

\newpage

\begin{figure}[tbp]
\begin{center}
\begin{minipage}{20\linewidth}
\epsfig{file=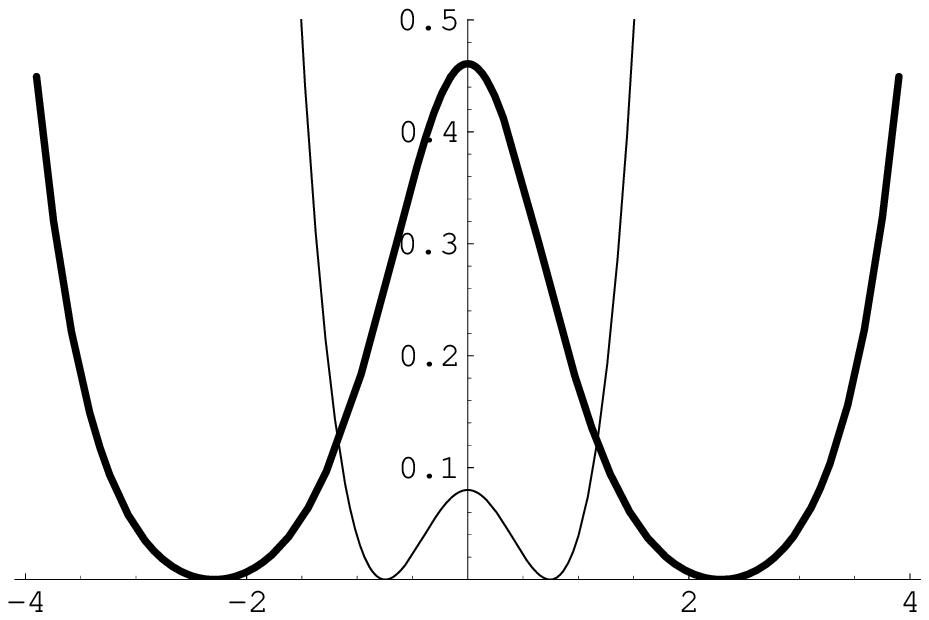}
\end{minipage}
\end{center}
\caption{The potential as a function of the scalar field
$\varphi$. A typical profile for $A$ significantly different of
zero (thin line) and when $A$ is close to zero.} \label{fig:fig1}
\end{figure}

\begin{figure}[tbp]
\begin{center}
\begin{minipage}{20\linewidth}
\epsfig{file=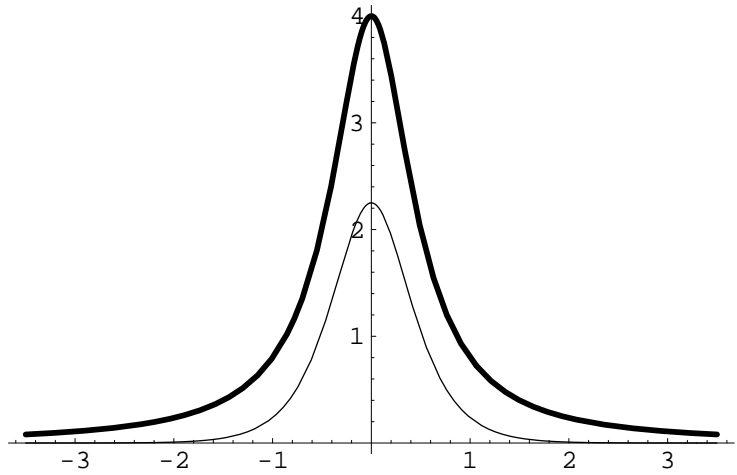}
\end{minipage}
\end{center}
\caption{The energy density dependence in $x$ when $B=2$ and
$A=0.5$ (thin line) and the vaccumless case.} \label{fig:fig2}
\end{figure}

\begin{figure}[tbp]
\begin{center}
\begin{minipage}{20\linewidth}
\epsfig{file=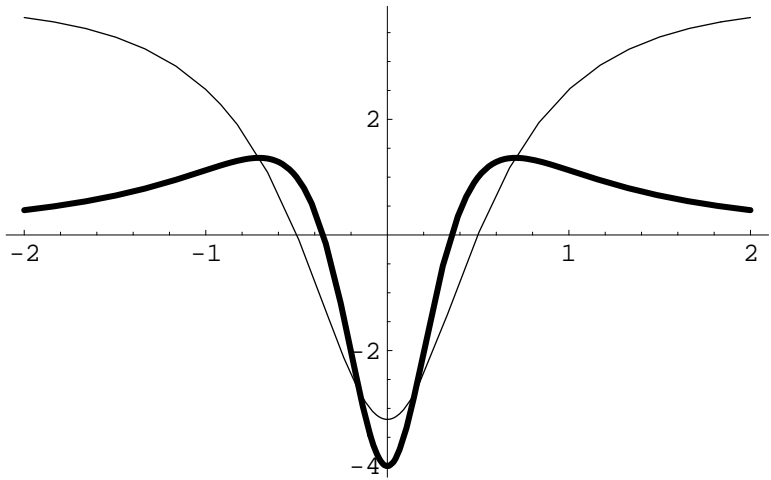}
\end{minipage}
\end{center}
\caption{The potential of the Scroedinger-like stability equation
as a function of the spatial variable, both in the case of
potential with degenerate vacua (thin line, $A=0.9$, $B=1$) as in
the vacuumless case.} \label{fig:fig3}
\end{figure}

\begin{figure}[tbp]
\begin{center}
\begin{minipage}{20\linewidth}
\epsfig{file=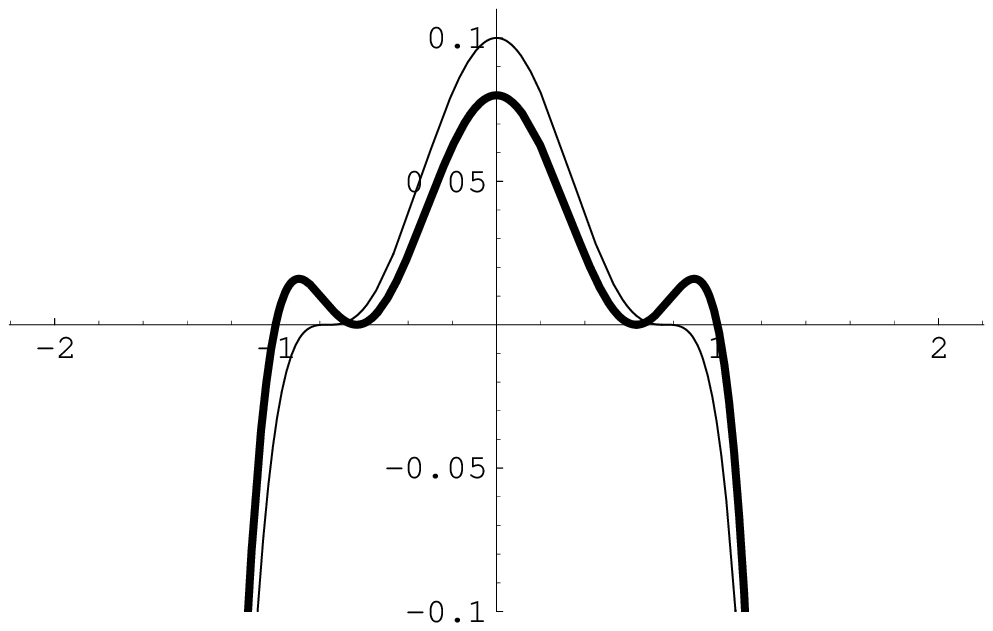}
\end{minipage}
\end{center}
\caption{The potential as a function of the scalar field
$\varphi$. A typical profile for $A$ significantly different of
zero and for the BLM model (thin line).} \label{fig:fig4}
\end{figure}


\bigskip

\newpage

\begin{thebibliography}{99}
\bibitem{jackiw}  E. D'Hoker and R. Jackiw, Phys. Rev. D \textbf{26} (1982)
3517; Phys. Rev. Lett. \textbf{50} (1982) 1719.

\bibitem{dhoker}  E. D'Hoker, D. Z. Freedman and R. Jackiw, Phys. Rev. D
\textbf{28} (1982) 2583.

\bibitem{menotti} P. Menotti and G. Vajente, Nucl. Phys. B {\bf
709} (2005) 465.

\bibitem{cho}  I. Cho and A. Vilenkin, Phys. Rev. D \textbf{59} (1999)
021701(R); \textbf{59} (1999) 063510.

\bibitem{dionisio99}  D. Bazeia, Phys. Rev. D \textbf{60} (1999) 067705.

\bibitem{sen} A. A. Sen, Int. J. Mod. Phys. D {\bf 10} (2001) 515.

\bibitem{deformed}  D. Bazeia, L. Losano and J. M. C. Malbouisson, Phys.
Rev. D \textbf{66} (2002) 101701(R).

\bibitem{BPS}  M. K. Prasad and C. M. Sommerfield, Phys. Rev. Lett. \textbf{%
35} (1975) 760; E. B. Bolgomol 'nyi, Sov. J. Nucl. Phys.
\textbf{24} (1976) 449.\newline

\bibitem{bazeia2}  D. Bazeia, R. F. Ribeiro and M. M. Santos, Phys. Rev. E
\textbf{54} (1996) 2943.

\bibitem{bazeiaPLA96}  D. Bazeia, M. M. Santos, Phys. Lett. A \textbf{217}
(1996) 28.

\bibitem{shiffman}  B. Chibisov and M. Shifman, Phys. Rev. D \textbf{56}
(1997) 7990.
\end{thebibliography}
\end{document}